\documentclass{emulateapj}
\usepackage{bm}
\usepackage{chngpage}
\usepackage{graphicx}
\usepackage{natbib}
\usepackage{mathrsfs}

\newcommand{\degree}{\ensuremath{^\circ}}
\def\apjl{ApJL}

\shorttitle{Pulsing White Dwarf}
\shortauthors{Geng et al.}

\begin{document}

\title{A MODEL OF WHITE DWARF PULSAR AR SCORPII}

\author{Jin-Jun Geng\altaffilmark{1,2}, Bing Zhang\altaffilmark{2,3,4}, Yong-Feng Huang\altaffilmark{1}}

\altaffiltext{1}{School of Astronomy and Space Science, Nanjing University, Nanjing 210046, China; gengjinjun@gmail.com, hyf@nju.edu.cn}
\altaffiltext{2}{Department of Physics and Astronomy, University of Nevada Las Vegas, NV 89154, USA; zhang@physics.unlv.edu}
\altaffiltext{3}{Department of Astronomy, School of Physics, Peking University, Beijing 100871, China}
\altaffiltext{4}{Kavli Institute of Astronomy and Astrophysics, Peking University, Beijing 100871, China}

\begin{abstract}
A 3.56-hour white dwarf (WD) - M dwarf (MD) close binary system, AR Scorpii, was recently reported to show 
pulsating emission in radio, IR, optical, and UV, with a 1.97-minute period, which suggests the
existence of a WD with a rotation period of 1.95 minutes. 
We propose a model to explain the temporal and spectral characteristics of the system.
The WD is a nearly perpendicular rotator, with both open field line beams sweeping the MD
stellar wind periodically. A bow shock propagating into the stellar wind accelerates electrons in the wind.
Synchrotron radiation of these shocked electrons can naturally account for the broad-band 
(from radio to X-rays) spectral energy distribution of the system. 
\end{abstract}


\section{INTRODUCTION}
A white dwarf (WD) - M dwarf (MD) binary, AR Scorpii (henceforth AR Sco), was recently reported
to emit pulsed broad-band (radio, IR, optical and UV) emission \citep{Marsh16}. The brightness of 
the system varies in long time scales with the orbital period of 3.56 hours, and shows pulsation
in short time scales with a period of 1.97 minutes. Interpreting the high pulsating frequency as the
``beat'' frequency of the system, the inferred rotation period of the WD is 1.95 minutes.
The spectral energy distribution of the pulsed emission supports a synchrotron origin for the 
radiation \citep{Marsh16}. These peculiar observational properties make AR Sco a unique system.
Although AR Sco could be in the evolutionary stage of the so-called 
intermediate polar \citep{Bookbinder87,Patterson94,Oruru12}, of which accretion is the main power 
source, the absence of accretion features in AR Sco demands another mechanism 
to explain the observations.

It has been suggested that if the dipole magnetic field of a WD is strong enough, it would behave
like a WD radio pulsar \citep{Zhang05}.\footnote{The so-called anomalous X-ray pulsars were
suggested as magnetized WDs \citep{Paczynski90,Usov93,Malheiro12,Lobato16,Mukhopadhyay16}, but they are now widely
accepted to be extremely magnetized neutron stars known as magnetars \citep{Duncan92,Thompson96}.
} The unique pulsating properties of AR Sco suggest that such WD pulsars indeed exist. Here 
we propose a WD-MD interaction model for the system. We show that 
the interaction between the WD pulsar open field line beams with the stellar wind
of MD naturally accounts for all the observational properties of the system.

\section{WHITE DWARF PULSAR}
White dwarfs are the final evolutionary state of stars whose masses are not large enough to 
become a neutron star \citep{Heger03}.
A group of WDs have a surface magnetic field ranging from $10^6-10^9$ G \citep{Wickramasinghe00}. 
Some of them spin with periods around one hour, possibly caused by the mass transfer from a 
companion star \citep{Ferrario97}.
These rapidly rotating magnetized WDs would mimic neutron star pulsars in many ways,
e.g., a co-rotating magnetosphere \citep{Goldreich69}, and possible pair production processes
\citep{Zhang05}.

The mass and radius of the WD in AR Sco are derived as
$M_{\rm WD} = 0.8 M_{\odot}$, $R_{\rm WD} = 7 \times 10^8$ cm, respectively \citep{Marsh16}.
With the measured period $P = 1.95~\rm{min}$ and period derivative 
$\dot{P} = 3.9 \times 10^{-13} \rm{s}~\rm{s}^{-1}$, the spin down luminosity of the WD pulsar
in AR Sco is $\dot{E}_{\rm rot} = - 4 \pi^2 I {\dot{P}}{P^{-3}}
\simeq 3.0 \times 10^{33} \rm{erg}~\rm{s}^{-1}$.
where $I = \frac{2}{5} M_{\rm WD} R_{\rm WD}^2$ is the moment of inertia of the WD.
The mean luminosity of AR Sco is $\simeq 1.7 \times 10^{33} \rm{erg}~\rm{s}^{-1}$
including the emission from the MD, and is $\simeq 1.6 \times 10^{31} \rm{erg}~\rm{s}^{-1}$ 
for the non-thermal emission only.
Therefore, the spin-down power of the WD can comfortably power the non-thermal 
radiation of AR Sco.

Assuming a magnetic dipole for the WD and considering a wind outflow from the magnetosphere,
the magnetic spin down power can be written as \citep[e.g.][]{Shapiro83,Xu01,Contopoulos06}
\begin{equation}
\dot{E}_{\rm mag} \simeq \frac{(2 \pi)^4 B_p^2 R_{\rm WD}^6}{6 c^3 P^4},
\end{equation}
where $B_p$ is the surface magnetic field, $c$ is the speed of light. The total spin down 
torque of the WD should be exerted from this dipole-wind component and a propeller
torque exerted from MD (corresponding to a spin down power of $\dot E_{\rm MD}$). 
In the following, we assume $\dot E_{\rm mag}
\gg \dot E_{MD}$. One can then derive the surface magnetic field at the polar cap,
\begin{eqnarray}
B_p & \simeq & \left( \frac{3 M_{\rm WD} c^3}{5 \pi^2 R_{\rm WD}^4 } P \dot{P} \right)^{1/2}  \nonumber \\
&=& 7.1 \times 10^8~\mathrm{G} \left(\frac{P}{1.95~\mathrm{min}} \frac{\dot{P}}{3.9 \times 10^{-13} \mathrm{s}~\mathrm{s}^{-1}}\right)^{1/2}. 
\end{eqnarray}
The light cylinder radius is $R_{\rm lc} = c P/2 \pi = 5.6 \times 10^{11} (P/1.95~\rm{min})$ cm,
which is greater than the distance between the WD and the MD, $d \sim 7.6 \times 10^{10} \rm{cm}$.
This suggests that the MD sits inside the magnetosphere of the WD, and significant interaction
between the WD wind and MD is expected.
The polar cap opening angle of the last open field line is \citep{Ruderman75}
\begin{equation}
\theta_{\rm open} =  \left( \frac{R_{\rm WD}}{R_{\rm lc}}\right)^{1/2} = 2 \degree \left( \frac{P}{1.95~\rm{min}}\right)^{-1/2},
\end{equation}
and the corresponding polar cap radius is 
\begin{equation}
R_{\rm pc} = R_{\rm WD}  \left( \frac{R_{\rm WD}}{R_{\rm lc}}\right)^{1/2} = 2.5 \times 10^7 \left( \frac{P}{1.95~\rm{min}}\right)^{-1/2} \rm{cm}. 
\end{equation}
The maximum available unipolar potential drop across the polar cap is
\begin{eqnarray}
\Phi_{\rm max} &=& \frac{2 \pi^2 B_p R_{\rm WD}^3}{c^2 P^2}  \simeq 3.9 \times 10^{11} \mathrm{statV}
\nonumber \\ 
& \times & 
\left(\frac{P}{1.95~\mathrm{min}} \right)^{-3/2} \left(\frac{\dot{P}}{3.9 \times 10^{-13} \mathrm{s}~\mathrm{s}^{-1}}\right)^{1/2},
\end{eqnarray}
which can accelerate electrons to a Lorentz factor
of $\gamma_e = q_e \Phi_{\rm max} /m_e c^2 \simeq 2.3 \times 10^8$,
where $q_e$ is the electron charge and $m_e$ is the electron mass. Pair production through
$\gamma-B$ and $\gamma-\gamma$ mechanisms is possible, so that the WD can act as
an active pulsar \citep{Zhang05,Kashiyama11}.

\section{GEOMETRY}
Before performing calculations on the radiation properties, we first infer the
geometry of AR Sco by analyzing the temporal characteristics of the pulses. 
Unlike radio pulsars which typically show a small duty cycle $<10\%$, AR Sco shows a large
duty cycle of about 50\% in the lightcurves in broad band. 
The emission likely comes from the MD rather than the WD itself \citep{Marsh16}. 
More interestingly, there are two peaks in each period, and each peak has exactly the same 
period. A natural explanation is that the WD is a nearly perpendicular rotator (the angle between the
spin and magnetic axes is close to $\alpha \sim 90 \degree$)\footnote{In our
calculations below, we adopt $\alpha \sim 90 \degree$ to simplify the problem.
In more realistic cases, one needs to consider a three-dimensional (3D) geometry.
In view that the MD is very close to the WD so that it opens up a large solid angle
to the WD, there is a wide range of $\alpha$ values (e.g. $\alpha > 45 \degree$)
that can make the two WD beams sweep the MD wind, and hence, to interpret the
data. Strictly speaking, $\alpha$ should not be too close to $90 \degree$, since
this would cause the eclipse of the shocked wind (see below) at certain orbital phases if the line-of-sight
is also in the orbital plane. A detailed geometry parameter search is beyond the scope
of this paper.} 
with both open field line beams sweeping the MD in each
rotation period. The orbital brightness modulation 
implies that the inclination of the orbital plane to the line-of-sight is small.
A deviation of the line-of-sight from the orbital plane would naturally produce the
uneven brightness of the two peaks in each spin cycle. 
All these clues lead to a special geometric configuration as shown in Figure 1. 
The spin axis of the WD points into the plane of the page, while the line-of-sight 
and the orbital plane are roughly on the page plane for the specific configuration in Fig. 1. 

The ratio between the durations of the active pulse and the quiescent time is $\sim$ 1:1 (duty cycle $\sim 50\%$).
This suggests that the opening angle of the entire radiation site seen from the WD should be $\sim 90\degree$.
Since the MD only opens a $2 R_{\rm MD}/d \sim 38 \degree$ angle from the WD,  
the actual location of the emission site should be at a larger radius from the MD center.
The interaction between the particle beam streaming out from the open field line regions and the MD wind
would lead to the formation of a bow shock, where the electrons
can be accelerated to give radiation.
In Fig. 1, when the open field lines of the WD are approaching the atmosphere of the MD,
there exists a position at which the ram pressure of the stellar wind balances the magnetic pressure.
We note this position as point ``A'' (its position will be calculated in Section 4.2)
and assume that radiation rises from this point.
In the realistic 3D case, this position should be a part of a ring-like region.
According to the duty cycle of the pulsation, the angle from the WD between point A and the MD center 
should be $\theta_{\rm A} \sim 90\degree/2 = 45\degree$.

\section{RADIATION}

\subsection{Synchrotron Radiation}

An outflow of relativistic particles from the open field line regions of the WD would impact the MD wind
and drive a bow shock into it. 
Since the magnetization of the WD wind is not known, there may or may not be a reverse shock.
In any case, since the accelerated particles are still within the magnetosphere of  
the WD, they would give rise to synchrotron radiation in the magnetic field of the magnetosphere.

The open field lines of the WD would be extruded by the ram pressure of the stellar wind
during the period when they sweep the MD wind.
When the polar cap faces the center of the MD, 
the effective opening angle of the open field lines reaches the maximum value (Fig. 1),
and the luminosity also reaches the peak value.
We define the head position of the bow shock at this epoch as point ``B''.
Since the magnetic pressure at B is larger than that at A, the magnetic pressure
would push point B to be near the surface of the MD.
Its distance to the center of the WD is $x \sim d-R_{\rm MD} \simeq 5.1 \times
10^{10}$~cm ($R_{\rm MD} = 2.5 \times 10^{10}$~cm is the radius of the MD)
and the corresponding magnetic field is $B_{\rm B} = B_p \left({x}/{R_{\rm WD}}\right)^{-3} 
\simeq 1860 ~\rm{G}$.  
Since the synchrotron emission spectral energy distribution (SED) we model is an average
one across different phases, we use an average magnetic field 
$\bar B = (B_{\rm A}+B_{\rm B})/2 \simeq 1200~\rm{G}$ ($B_{\rm A}$ is
the magnetic field at A and is obtained according to its position, see Section 4.2) 
in our calculations.

For a relativistic electron of Lorentz factor $\gamma_e$, its radiation power is \citep{Rybicki79}
\begin{equation}
\dot{\epsilon} = \gamma_e^2 \sigma_{\rm T} \bar{B}^2 c / 6 \pi = 1.5 \times 10^{-9} \gamma_{e}^2~\rm{erg}~\rm{s}^{-1},
\end{equation}
where $\sigma_{\rm T}$ is the Thomson cross section.
Hereafter the convention $Q_x = (Q/10^x)$ is adopted in cgs units.
Its cooling time scale can be calculated as $t_{\rm cool} = \gamma_e m_e c^2/\dot{\epsilon}$.
It is reasonable to assume that the relativistic electrons that give rise to synchrotron radiation 
obey a broken-power-law distribution, i.e., 
\begin{equation}
\frac{dn_e}{d\gamma_e}=\left\{
\begin{array}{ll}
C \gamma_{e}^{-p},&\gamma_m\leq\gamma_e\leq\gamma_c,\\
C \gamma_c \gamma_{e}^{-p-1},&\gamma_c<\gamma_e\leq\gamma_{\rm max},
\end{array}\right.
\label{eq:electron-dist}
\end{equation}
where $C \simeq (p-1) \gamma_m^{p-1} n_e$, 
$\gamma_m$ is the typical Lorentz factor, $\gamma_c$ is the cooling Lorentz factor,
$\gamma_{\rm max}$ is the maximum Lorentz factor, and $n_e$ is the number density of the electrons.
Then the peak frequency of the corresponding synchrotron spectrum is 
\begin{equation}
\nu_m = \frac{3 x_p q_e \bar{B} \gamma_m^2}{4 \pi m_e c} = 2.5 \times 10^{9} \gamma_m^2~\rm{Hz},
\end{equation}
where $x_p \simeq 0.5$ \citep{Wijers99}.
Inspecting the SED (\citealt{Marsh16}, Fig. 2), we find $\nu_m \simeq 5 \times 10^{12}~\rm{Hz}$, 
which gives $\gamma_m = 45$.
The cooling Lorentz factor $\gamma_c$ can be estimated by equating $t_{\rm cool}(\gamma_c)$ with
the mean dynamical time of the shock $t_{\rm dyn} \simeq \frac{45\degree/2}{360\degree} P$, where 
$45\degree$ is the angle for the WD beam to sweep through before reaching the peak, and the half
value reflects the mean angle, which defines the mean dynamical time scale.
We then obtain $\gamma_c = 73~ (P/1.95~\rm{min})^{-1}$.
The Lamor radius of a relativistic electron is $R_{L} = {\gamma_e m_e c^2}/{q_e \bar{B}}$,
and its acceleration time scale can be calculated as $t_{\rm acc} \simeq R_{L}/c$.
Equating $t_{\rm acc} (\gamma_{\rm max})$ with $t_{\rm cool} (\gamma_{\rm max})$,
one can obtain the maximum Lorentz factor of the shocked electrons as
$\gamma_{\rm max} = \left( {6 \pi q_e}/{\sigma_T \bar{B}}\right)^{1/2} = 3.4 \times 10^6$.
The corresponding maximum Lamor radius is
$R_{L, \rm{max}} = 4.8 \times 10^6~\rm{cm}$.

Assuming that the width of the emission shell is $\eta R_{\rm MD}$, 
the shell volume $V$ is $2 \pi (1-\cos(\theta_{\rm B})) \eta R_{\rm MD}^3$,
where $\theta_{\rm B}$ is the half opening angle of the shell seen from the MD,
and $\theta_{\rm B} \simeq 90\degree-\arcsin(R_{\rm MD}/d) = 71 \degree$ (Fig. 1).
The value of $\eta$ may be inferred from the Lamor radius of the most energetic electron, i.e.,
$\eta \simeq 2 R_{L,\rm{max}} / R_{\rm MD} = 4 \times 10^{-4}$. Since $\nu_m < \nu_c$,
the synchrotron emission is in the slow cooling regime, and the peak flux density is at $\nu_m$, which reads
\begin{equation}
F_{\nu,\rm{peak}} = \frac{\sqrt{3}q_e^3 \bar{B}}{m_e c^2} \frac{n_e V}{4 \pi D_L^2},
\end{equation}
where $D_L = 116~\rm{pc}$ is the distance of AR Sco to the observer.
Observationally, the spectrum of AR Sco shows a peak flux density  
$\simeq 1.6 \times 10^{-24}~\rm{erg}~\rm{cm}^{-2}~\rm{s}^{-1}~\rm{Hz}^{-1}$
\citep{Marsh16}. We therefore obtain
\begin{eqnarray}
n_e & = & 3.5 \times 10^{8}  \left(\frac{F_{\nu,\rm{peak}}}{1.6 \times 10^{-24}~\rm{erg}~\rm{cm}^{-2}~\rm{s}^{-1}~\rm{Hz}^{-1}} \right) \nonumber  \\
& \times & \left( \frac{\eta}{4 \times 10^{-4}}\right)^{-1} \rm{cm}^{-3}.
\label{eq:ne}
\end{eqnarray}
This gives an estimate on the parameter $n_e$ of the emission region. 
The source of the emitting electrons remains not specified so far, which we 
will constrain next.

\subsection{Source of Electrons}

The \cite{Goldreich69} charge number density of the WD pulsar at position B is estimated as
\begin{equation}
n_{\rm GJ} = \frac{\mathbf{\Omega} \cdot \mathbf{B}_{\rm B}}{2 \pi q_e c} = 1.1~\rm{cm}^{-3}.
\end{equation}
This is significantly lower than the required $n_e$ in Eq.(\ref{eq:ne}).
The primary electrons ejected from the polar cap may produce secondary 
electron-positron pairs through some pair production process \citep[e.g.][]{Zhang00}.
However, the pair multiplicity $\chi = n^{\pm} / n_{\rm GJ}$ is at most $10^6$ even for young radio 
pulsars \citep[e.g.][]{Arons09}. This is not enough to account for Eq.(\ref{eq:ne}), which demands
$\chi \sim 10^8$. We therefore conclude that the synchrotron emitting electrons are not from
the WD wind.

We now consider the possibility that the emitting electrons come from the stellar wind of the MD.
Assuming that the wind velocity $v$ is two times of the escape velocity from the MD 
(stellar wind velocities are generally not much greater than the surface escape
velocity, e.g., \citealt{Wood04}), i.e.,
\begin{equation}
v = 2 v_{\rm esc} = 2 \sqrt{\frac{2 G M_{\rm MD}}{R_{\rm MD}}} = 1.1 \times 10^{8} \rm{cm}~\rm{s}^{-1},
\end{equation}
the corresponding mass-loss rate of this stellar wind would be
$\dot{M} = 4 \pi R_{\rm MD}^2 v  n_e m_p /\zeta = 4.1 \times 10^{-11} (\zeta/0.2)^{-1} M_{\odot}/{\rm yr}$,
where $\zeta$ denotes the fraction of electrons that are accelerated and
$n_e$ has been taken as $3.5 \times 10^8$ cm$^{-3}$ as required in Eq. (10).
This value is within the range of mass-loss rate of MDs in the literature,
$10^{-15} - 10^{-10} M_{\odot}/{\rm yr}$ \citep[e.g.][]{Mullan92,Wood01,Vidotto14}.
It indicates that the wind from the MD could be a reasonable source for electrons.

The position of the balance point A (see Fig. 1 and Section 3), can be found using the
balance between the ram pressure of the wind and the magnetic pressure.
We define the distance between A and the MD center as $r_{\rm A}$. The 
distance between A and the WD center may be roughly estimated as $d$.
The relative velocity between the wind and the magnetic field lines is thus
$v_{\rm rel} =  \frac{2 \pi}{P} d$.
The condition that the magnetic pressure balances the wind ram pressure gives
\begin{equation}
\frac{B_{\rm A}^2}{8 \pi} = \left(\frac{r_{\rm A}}{R_{\rm MD}}\right)^{-2} \frac{1}{\zeta} n_e m_p v_{\rm rel}^2,
\label{eq:pressures}
\end{equation} 
where $B_{\rm A} = B_p \left({d}/{R_{\rm WD}}\right)^{-3}$ is the magnetic field strength
at point A, $m_p$ is the proton mass, and the factor $(\frac{r_{\rm A}}{R_{\rm MD}})^{-2}$
takes into account the correction of the electron density $n_e$ from point B to point A
due to the radial expansion of the wind.
Using equations (\ref{eq:ne}) and (\ref{eq:pressures}), one obtains
\begin{eqnarray}
r_{\rm A} & = & 5.0 \times 10^{10} \left(\frac{F_{\nu,\rm{peak}}}{1.6 \times 10^{-24}~\rm{erg}~\rm{cm}^{-2}~\rm{s}^{-1}~\rm{Hz}^{-1}} \right)^{1/2}  \nonumber \\
& \times & \left( \frac{\eta}{4 \times 10^{-4}}\right)^{-1/2}  \left( \frac{\zeta}{0.2}\right)^{-1/2}  ~\rm{cm}.
\end{eqnarray}
Another way to derive $r_{\rm A}$ is from the geometric relation in Fig. 1.
For the isosceles triangle defined by point A and the centers of the two stars, 
one has $\sin \theta_{\rm A} / r_{\rm A} = \sin(90\degree-\theta_{\rm A}/2) / d$,
which gives $r_{\rm A} = 5.8 \times 10^{10}~\rm{cm}$ for $\theta_{\rm A} \sim 45 \degree$.
One can see that the values of $r_{\rm A}$ are consistent with each other using two methods,
suggesting that the emitting electrons are indeed from the MD wind.

\subsection{The Spectrum}

Using the electron distribution in Equation (\ref{eq:electron-dist}), one can calculate the synchrotron spectrum.
The spectrum is characterized by a broken power law separated by three frequencies,
the self-absorption frequency $\nu_a$, the injection frequency $\nu_m$, and the cooling frequency $\nu_c$
\citep{Sari98}. The characteristic frequencies
$\nu_m$ and $\nu_c$ are derived from $\gamma_m$ and $\gamma_c$, respectively.
The spectrum in the spectral regime $\nu_m < \nu < \nu_c$ has $\nu F_{\nu} \propto \nu^{-(p-3)/2}$.
According to the observed SED of AR Sco \citep{Marsh16}, the source was also detected in the X-ray 
band by {\em Swift} X-Ray Telescope (even though not pulsed). The X-ray data constrains the value of $p$
to be around 2.4.
The self-absorption coefficient at frequency $\nu$ can be calculated as (e.g., \citealt{Rybicki79,Wu03})
\begin{equation}
\kappa_{\nu} \simeq c_p \frac{q_e}{\bar{B}} n_e \gamma_m^{-5} \left( \frac{\nu}{\nu_m}\right)^{-5/3},
\end{equation}
where $c_p =  \frac{8 \pi^2}{9~2^{1/3} \Gamma(1/3)} \frac{p+2}{p+2/3} (p-1) \simeq 5.2$, and 
$\Gamma(1/3)$ is the Gamma function of argument $1/3$.
The self-absorption frequency can be solved when the optical depth $\kappa_{\nu_a} \eta R_{\rm MD}$ equals 1,
which gives
\begin{eqnarray}
\left( \frac{\nu_a}{\nu_m}\right)^{5/3} & = & 4.2 \times 10^{-5} \left(\frac{F_{\nu,\rm{peak}}}{1.6 \times 10^{-24}~\rm{erg}~\rm{cm}^{-2}~\rm{s}^{-1}~\rm{Hz}^{-1}} \right) \nonumber \\
& \times & \left( \frac{\nu_m}{5 \times 10^{12}~\rm{Hz}}\right)^{-5/2}.
\end{eqnarray}
One gets $\nu_a \simeq 2.4 \times 10^{-3}~\nu_m$ for AR Sco.
Figure 2 displays the comparison between our analytical model SED and the observed
SED \citep{Marsh16}. One can see that the model can well interpret the data. One interesting
feature is that the observed flux at the thermal peak exceeds the model spectrum of the 
MD thermal emission. However, adding the contribution of the non-thermal synchrotron
component (which extends all the way to the X-ray band), the total model flux matches
the observations very well.

\section{Summary and Discussion}

We have shown that the peculiar observations of the pulsating AR Sco system can be
understood with the framework of interaction between the WD pulsar's open field line beams
and the wind of the MD. 
The observational data demand a nearly perpendicular rotator for the WD pulsar, and a 
near edge-on orbital configuration for the observer on Earth. In order to interpret the
observed SED, the required electron number density is too high for a WD wind. Rather
electrons accelerated by a bow shock into the MD wind can produce the right amount of
electrons to interpret both the shape and the normalization of the SED. 

In our model, although the magnetic field lines of the WD are likely ordered, an observer
sees a hemisphere where magnetic field lines have different directions so that on average
the directional information cancels out (e.g. in Fig. 2, the field lines above and
below point B have opposite orientations). One would therefore do not expect significant
circular polarization \citep{Matsumiya03},
which is consistent with the observations \citep{Marsh16}.

Our model suggests that rapidly rotating, highly magnetized WDs can indeed
behave like radio pulsars, as has been speculated in the past \citep{Zhang05}.
The rarity of these WD pulsars \citep{Kepler13} may be due to the conditions to produce an active
magnetosphere via pair production are much more stringent for WDs than NSs.
The peculiarity of AR Sco lies in its extremely short period and its close proximity 
with its MD companion. According to our modeling, the observed emission is from the
shocked MD wind rather than from the WD pulsar itself. However, if some WDs indeed behave
as pulsars, one would expect to directly detect emission from WD pulsars in the 
future. GCRT J1745-3009 might be another, less energetic, transient WD pulsar
\citep{Zhang05} at a distance beyond 1 kpc from the earth \citep{Kaplan08}.

\acknowledgments
We thank the anonymous referee for valuable suggestions, and Yuan-Pei Yang
for helpful discussion on this paper.
This work is partially supported by 
the National Basic Research Program of China with Grant No. 2014CB845800, 
and by the National Natural Science Foundation of China (grant Nos. 11473012 and 11303013).
J.J.G. acknowledges the China Scholarship Program to conduct research at UNLV.

\begin{figure}
   \begin{center}
   \includegraphics[scale=0.5]{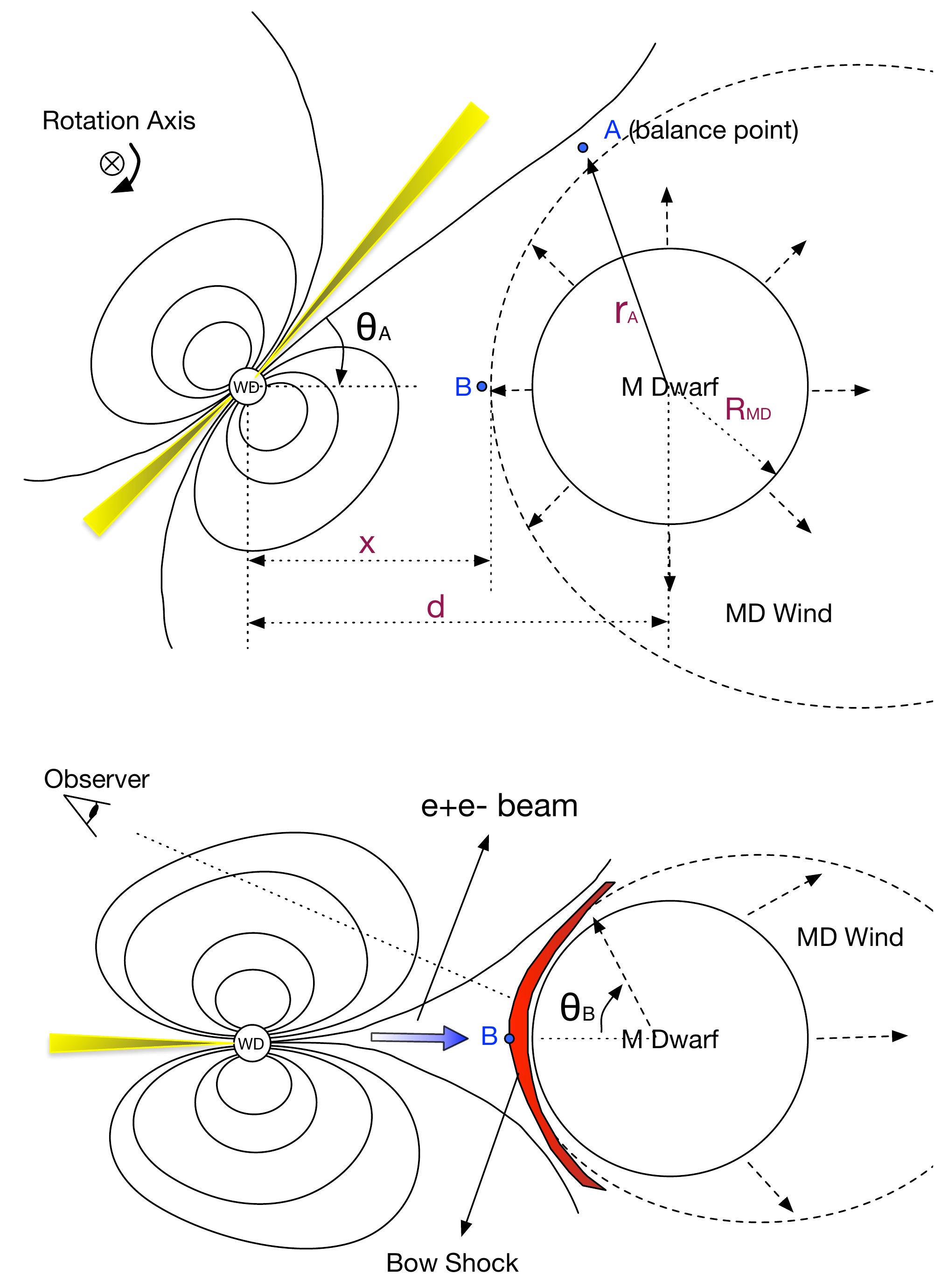}
   \caption{Schematic pictures of the WD/MD binary in AR Sco. 
   The upper panel shows the episode that the emission begins to rise when
   the open field lines approach the balance point A (see Section 3).
   The lower panel shows the episode when the emission reaches the peak emissivity.
   Note that the rotation axis of the WD points into the the page plane, 
   while the orbital plane and the line-of-sight are all roughly within the page plane.
   }
   \label{Fig:plot1}
   \end{center}
\end{figure}

\begin{figure}
   \begin{center}
   \includegraphics[scale=0.5]{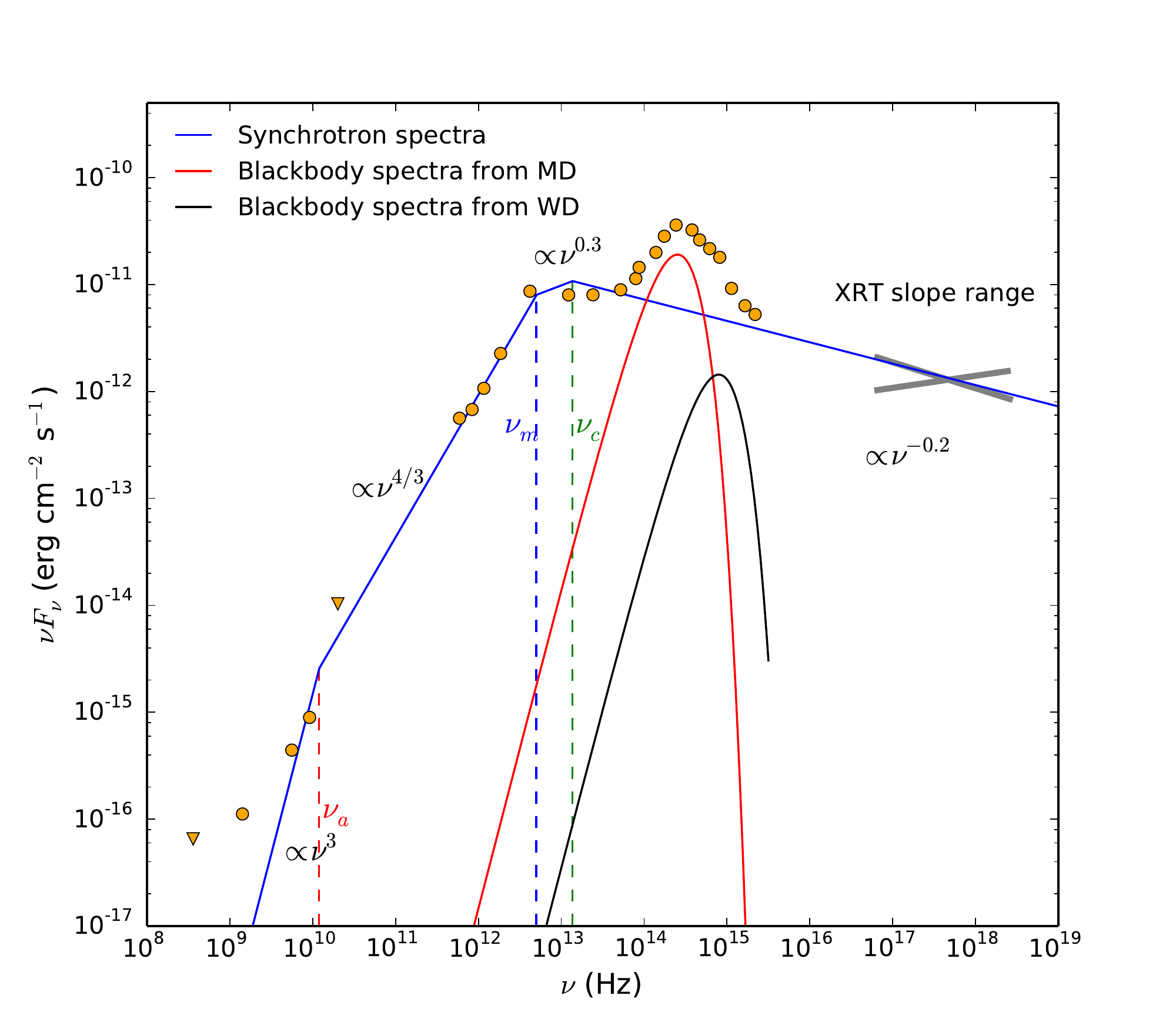}
   \caption{The observed SED of AR Sco compared with our analytical model results.
   The red and black solid line represent the blackbody spectrum from the MD (temperature of 3100~K)
   and the WD (temperature of 9750~K), respectively. 
   The blue solid line is the analytical model synchrotron spectrum.
   The observational data points (without error bars) are obtained from Figure 4 of \cite{Marsh16}.
   }
   \label{Fig:plot2}
   \end{center}
\end{figure}

\end{document}